\documentclass[pra,twocolumn]{revtex4-1}

\usepackage{amsmath}
\usepackage{graphicx}
\usepackage{microtype}
\usepackage[usenames]{color}
\usepackage{hyperref}

\usepackage{soul}
\usepackage[normalem]{ulem}

\newcommand{\beq}{\begin{equation}}
\newcommand{\eeq}{\end{equation}} 
\newcommand{\beqa}{\begin{eqnarray}}
\newcommand{\eeqa}{\end{eqnarray}}
\newcommand{\ba}{\begin{array}}
\newcommand{\ea}{\end{array}}

\begin{document}

\title{Unitary Fermi superfluid near the critical temperature:\\
thermodynamics and sound modes from elementary excitations}
\author{G.~Bighin$^{1}$, A.~Cappellaro$^{2}$, and L.~Salasnich$^{3,4,5}$} 
\affiliation{$^{1}$Institut fur Theoretische Physik, Universit\"at Heidelberg,
Philosophenweg 19, D-69120 Heidelberg, Germany\\
$^{2}$Institute of Science and Technology Austria (ISTA), 
Am Campus 1, 3400 Klosterneuburg, Austria \\
$^{3}$Dipartimento di Fisica ``Galileo Galilei'' and QTech, 
Universit\`a di Padova, Via Marzolo 8, 35122 Padova, Italy\\
$^{4}$Istituto Nazionale di Fisica Nucleare, Sezione di Padova, 
via Marzolo 8, 35131 Padova, Italy\\
$^{5}$Istituto Nazionale di Ottica del Consiglio Nazionale 
delle Ricerche, via Carrara 2, 50019 Sesto Fiorentino, Italy}

\begin{abstract} 
We compare recent experimental results [Science {\bf 375}, 528 (2022)] 
of the superfluid unitary Fermi gas near the critical temperature with a 
thermodynamic model based on the
elementary excitations of the system. 
We find good agreement between experimental data and our 
theory for several quantities such as first sound, second sound, 
and superfluid fraction. We also show that mode mixing between 
first and second sound occurs. Finally, we characterize the response 
amplitude to a density perturbation: close to the critical temperature both 
first and second sound can be excited through a density perturbation, 
whereas at lower temperatures only the first sound mode exhibits a 
significant response. 
\end{abstract} 

\maketitle

\section{Introduction}

The unitary Fermi gas, i.e.~a gas of resonantly interacting fermions in the 
limit for which the scattering length diverges, constitutes a fundamental 
model in many-body physics \cite{Zwerger:2011vb,Randeria:2012kq}, and it has 
been the subject of a great deal of theoretical 
\cite{Hu:2007tv,Enss:2011gk,Werner:2006uh,Zwerger:2016tc} and experimental 
investigations \cite{Thomas:2005wy,Nascimbene:2010wo,Sagi:2012wg,
Ku:2012ue,Patel:2020th,Li:2022vn}. It is a unifying paradigm, of remarkable 
importance for several different subfield of physics, from ultracold quantum 
gases, nuclear matter up to high-energy physics. 
Indeed, the unitary Fermi gases is the non-relativistic setup which appears to
be closer to the perfect fluidity as conjectured by string-theoretical arguments
\cite{Schafer:2009vf,Schafer:2014tp},
i.e.~a fluid saturating the lower bound on the shear viscosity-entropy ratio \cite{Kovtun:2005tz}.

The scale invariance of the system means that, as the scattering length 
diverges, the only energy scale in the system at $T=0$ is the Fermi energy $T_F$
and that all thermodynamic and transport quantities can be expressed 
as universal functions,
depending on $T/T_F$ only.  As a consequence, the unitary Fermi gas has emerged as  
standard testbed for several different many-body theoretical approaches 
\cite{Zwerger:2016tc}. 
A remarkable possibility for studying the unitary Fermi gas comes from 
ultracold fermions in the vicinity of a Feshbach resonance: as an external 
magnetic field is tuned across the resonance the fermion-fermion interaction 
can assume all values from weakly to strongly attractive -- in a scenario 
known as the BCS-BEC crossover. As a consequence, the system varies with 
continuity from the BCS limit where fermions form large Cooper pairs 
over a definite Fermi surface, to the BEC limit where fermions form 
tightly-bound bosonic molecules. Critically, the unitary Fermi gas is 
to be found between these two limits, so that its superfluid transition does 
not simply correspond to the usual BCS or BEC paradigms, rather being 
due to a delicate interplay between the two \cite{Randeria:2012kq}.

Through the years, it has been shown that this interplay can be described 
within a thermodynamic approach \cite{Bulgac:2006tu,Bulgac:2007tc,
Salasnich:2010jw,Salasnich:2011tj,Bighin:2015tj,Bighin:2016ty,Bighin:2017ug} 
including temperature-independent 
single-particle and collective elementary excitations of 
the unitary Fermi gas. Such an approach describes with great precision 
a number of features, with favorable comparisons with experimental data 
\cite{Salasnich:2010jw,Salasnich:2011tj,Bighin:2015tj}. Moreover, it has 
been demonstrated that this approach, originally proposed by Landau on 
phenomenological grounds \cite{Landau:1980ue}, can be justified via
beyond-mean-field treatments of a Fermi gas, such as the 
Nozi\`eres-Schmitt-Rink (NSR) \cite{Nozieres:1985zz} and the Gaussian 
pair-fluctuations approach (GPF) \cite{Hu:2007ez,klimin-2011,klimin-2012,Tempere:2012tt}, 
in which a systematic treatment of the order parameter and its fluctuations 
leads to a rigorous \textit{ab initio} theory with essentially the same physical 
content: BCS-like single-particle excitations and collective excitations 
with a Bogoliubov-like dispersion. It is also important noting that 
it has been recently pointed out that beyond-GPF corrections are quite small 
in the broken symmetry phase \cite{Mulkerin:2016vs,Mulkerin:2022uw}.

In such a complex scenario, it is fundamental to identify a diagnostic
tool allowing for a comparison between theory and experiment.
From this perspective, sound propagation is certainly a promising candidate 
for a variety of reasons. On a conceptual standpoint it can be derived on a 
hydrodynamic basis by connecting thermodynamic and transport 
quantities within the framework of Landau two-fluid theory 
\cite{Landau:1941vk,Landau:1980ue}, with no need  -- in principle -- 
to refer to the particular features of the microscopic 
constituents. From an experimental perspective, it has been recently shown
that both modes predicted by the above mentioned Landau theory 
can be excited by a density-perturbing protocol driven by external laser fields 
\cite{Hoffmann:2021wi}. 

Along this path, the most recent experimental breakthroughs concerning the
unitary  Fermi gas \cite{Patel:2020th,Li:2022vn} allowed for the measurement
of many  properties at unprecedented level of precision, providing very
stringent  benchmarks for the theoretical models. The present paper
demonstrates that  a thermodynamic theory accounting 
for temperature-independent elementary
single-particle and  collective excitation is able to reproduce with excellent
precision the most recent measurements on the sound velocity. In particular, 
for first sound, second sound, and superfluid fraction we find very good 
agreement between experimental data \cite{Li:2022vn} and our theory, 
taking into account the mode mixing between first and second sound. 
We also prove that around the critical temperature both the first and 
second sound modes may be detected with a density perturbation, but only 
the first sound mode has a significant density response at very 
low temperatures. 

\section{Describing the unitary Fermi gas from elementary 
excitations} 

Following an approach pioneered by Landau \cite{Landau:1980ue}, we describe 
the low-temperature thermodynamics
of a uniform unitary Fermi gas, consisting of N particles contained in a 
volume $V=L^3$, in the superfluid phase, 
by means of its temperature-independent single-particle BCS-like excitations 
and collective Bogoliubov-like excitations. 
Within this framework, an effective Hamiltonian 
describing the system can be written down \cite{Salasnich:2010jw} as
\beq 
{\hat H} =  {3\over 5} \xi  \epsilon_F N + \sum_{\sigma=\uparrow,\downarrow} 
\sum_{\bf p} \epsilon_\text{sp}(p) \ {\hat c}_{{\bf p}\sigma}^{\dagger} 
{\hat c}_{{\bf p}\sigma} + \sum_{\bf q} \omega_\text{col}(q) \ 
{\hat b}_{\bf q}^{\dagger} {\hat b}_{\bf q} \; , 
\label{hamilt}
\eeq
where the ${\hat c}_{{\bf p}\sigma}^{\dagger}$ 
(${\hat c}_{{\bf p}\sigma}$) operator creates (annihilates) a single-particle 
excitation, respectively, with linear momentum $\textbf{p}$, spin $\sigma$, 
and energy $\epsilon_\text{sp}(p)$, whereas the ${\hat b}_{\bf p}^{\dagger}$
(${\hat b}_{\bf p}$) operator creates (annihilates) a bosonic collective 
excitation, respectively, of linear momentum ${\bf q}$ and energy 
$\omega_\text{col}(q)$.

The first term of Eq.~(\ref{hamilt}) represents the ground-state energy of 
the uniform unitary Fermi gas \cite{Chen:2005hj,Levin:2010vb}, $\xi$ being 
the celebrated Bertsch parameter $\xi \simeq 0.38$ \cite{bertsch} having 
also introduced the Fermi energy $\epsilon_F= \hbar^2(3\pi^2 n)^{2/3}/(2m)$ 
of a non-interacting Fermi gas of density $n=N/V$. 

The second and third terms represent the contribution from off-condensate 
fermionic single-particle excitations and collective modes, respectively. 
Of course these terms do not have any use until the dispersions of the
temperature-independent elementary 
excitations are specified. In Refs.~\cite{Salasnich:2008fr,Salasnich:2010ws} 
the dispersion relation of collective elementary excitations 
has been derived as
\beq
\omega_\text{col}(q) = \sqrt{\frac{q^2}{2m} \bigg( 2 m c_B^2 + 
{\frac{\lambda}{2m}}q^2\bigg) } \; , 
\eeq
where $c_B = \sqrt{\xi/3}\ v_F$ is the Bogoliubov sound velocity with 
$v_F=\sqrt{2\epsilon_F/m}$ the 
Fermi velocity of a non-interacting Fermi gas. 
Here, we set  $\lambda=0.02$, by fitting the spectrum of bosonic
collective modes obtained from the GPF theory \cite{Bighin:2015tj}
(see \cite{Hu:2007ez,Tempere:2012tt,marini} for an exhaustive review on the
basics of this approach).

However, the collective modes correctly describe only 
the low-energy density oscillations of the system;
at higher energies one expects the appearance of 
fermionic single-particle excitations starting 
from the threshold above which 
 Cooper pairs break down \cite{Chen:2005hj,Bulgac:2006tu,Bulgac:2007tc,
Bulgac:2008uy,Magierski:2009ut}. The dispersion of these
temperature-independent single-particle 
elementary excitations can be written as 
\beq 
\epsilon_\text{sp}(p) = 
\sqrt{\big({p^2\over 2m} - \zeta \epsilon_F \big)^2 + \Delta_0^2}  
\eeq
where $\zeta$ is a parameter taking into account the interaction 
between fermions and the reconstruction of the Fermi surface 
close to the critical temperature ($\zeta \simeq 0.9$ according to 
 accurate Monte Carlo results \cite{Magierski:2009ut}). 
Moreover, $\Delta_0$ is the gap parameter,
with $2\Delta_0$ the minimal energy to break a Cooper 
pair \cite{Chen:2005hj}. 
Notice that the gap energy $\Delta_0$ of the unitary Fermi gas 
at zero-temperature has been calculated with Monte Carlo 
simulations \cite{Magierski:2009ut,Carlson:2003vd,
Chang:2004wf,Carlson:2005tt} and found to be 
$\Delta_0=\gamma \epsilon_F $, with $\gamma \simeq 0.45$.
Let us also notice that, while $\Delta$ certainly has a temperature
dependence, the inclusion of a phenomenological thermal profile 
(as proposed, for instance in \cite{prozorov}) in our framework does not produce
any significant change in the sound velocities and the superfluid fraction.

\section{Universal thermodynamics at unitarity}

The Helmholtz free energy $F$ of the system is given by the usual formula 
$F = -k_B T \ln{\mathcal{Z}}$,
where we introduced the partition function $\mathcal{Z}$ of the 
system \cite{Altland:2006}, defined as
\beq 
\mathcal{Z} = \mathrm{Tr} [e^{-{\hat{H}}/k_B T}] \; .
\eeq
Similarly to Eq.~(\ref{hamilt}), the free energy of the unitary Fermi gas  
can be written as $F = F_0+F_\text{col}+F_\text{sp}$, 
where $F_0$ is the free energy of the ground-state,
\beq
F_{\text{sp}} = - \frac{2}{\beta} \sum_{\mathbf{k}} 
\ln [ 1 + e^{-\beta E_k} ]
\eeq
is the free energy of fermionic single-particle excitations and finally
\beq
F_{\text{col}} = - \frac{1}{\beta} \sum_{\mathbf{q}} 
\ln [ 1 - e^{-\beta \omega_q} ]
\eeq
is the free energy of the bosonic collective excitations.

As discussed in detail in Ref.~\cite{Salasnich:2010jw}, 
the total Helmholtz free energy $F$ of a unitary Fermi gas
in the superfluid phase can be then written as 
\beq 
F = N \epsilon_F \Phi (x) \; , 
\label{free} 
\eeq 
where, due to the scale-invariance of the system, $\Phi(x)$ is a function 
of the scaled temperature $x \equiv T/T_F$ only, having defined the 
Fermi temperature $T_F=\epsilon_F/k_B$. Explicitly, $\Phi(x)$ takes the 
following form 
\beqa 
\Phi(x) &=& {3\over 5}\xi - 3 x \int_0^{+\infty} 
\ln{\left[ 1 + e^{-{\tilde \epsilon}_\text{sp}(u)/x}\right]} u^2 \mathrm{d} u
\nonumber
\\
&+& {3\over 2} x \int_0^{+\infty} 
\ln{\left[ 1 - e^{-{\tilde \omega}_\text{col}(u)/x} \right]}u^2 
\mathrm{d} u \; . 
\label{free-scaled} 
\eeqa
Note that the discrete summations have been replaced by integrals, 
and that we set ${\tilde \omega}_\text{col}(u)=\sqrt{u^2(4\xi/3 + 
\lambda u^2)}$ and ${\tilde \epsilon}_\text{sp}(u)
=\sqrt{(u^2-\zeta)^2+\gamma^2}$. 

We now aim at calculating the thermodynamics of the system in terms of 
the universal function $\Phi(x)$ and its derivatives. We start from
the entropy $S$, which is readily calculated from the free energy $F$ 
through the relation
\beq 
S = - \left({\frac{\partial F}{\partial T}}\right)_{N,V} \; , 
\eeq
from which we immediately get
\beq 
S = - N k_B \Phi'(x) \; .    
\label{entropy}
\eeq 
where $\Phi'(x)$ is the first derivative of $\Phi$ with respect to $x$.
Furthermore, the internal energy $E = F + T S$,
can immediately be rewritten as 
\beq 
E = N \epsilon_F 
\left[ \Phi (x) - x \ \Phi' (x) \right] \;  
\label{internal}
\eeq
and, similarly, the pressure $P$ is related to  the free energy $F$ by
the simple relation 
\beq 
P = - \left( {\partial F\over \partial V} \right)_{N,T} \; ,
\eeq
which we now rewrite in terms of $\Phi(x)$ and its derivatives as
\beq 
P = \frac{2}{3} n \epsilon_F 
\left[\Phi (x) - x 
\Phi'(x) \right] \; .
\label{pressure}
\eeq
As a consistency check of our simple analytical model, let us underline that, by combining 
Eq. \eqref{internal} and Eq. \eqref{pressure} one can easily recover the well-known relation
$PV = (2/3) E$ for unitary fermions \cite{Thomas:2005wy}.

\section{Superfluid fraction and critical temperature} 

According to Landau's two fluid theory \cite{Landau:1941vk,Landau:1980ue}, the total number density $n$ of a 
system in the superfluid phase can be written as
\beq 
n = n_\text{s} + n_\text{n} \; , 
\eeq
where $n_\text{s}$ is the superfluid density and $n_\text{n}$ is the normal 
density \cite{Landau:1980ue}. Naturally, at zero temperature the whole system 
is in the superfluid phase, and one has $n_\text{n}=0$ and $n=n_\text{s}$. 
As the temperatures increases, the normal density $n_\text{n}$ increases, 
as well, until at the critical temperature $T_c$ one has $n_\text{n}=n$ and, 
correspondingly, $n_\text{s}=0$. Within our scheme, the normal density of a 
unitary gas is given the sum of two contributions 
\beq 
n_\text{n} = n_\text{n,sp} + n_\text{n,sp} \; ,  
\eeq
i.e.~a contribution $n_\text{n,sp}$ from to the single-particle excitations 
and a contribution $n_\text{n,col}$ from collective excitations. We note that 
in the BCS limit of the BCS-BEC crossover one expects $n_\text{n,sp}$
to be the dominating contribution, whereas in the BEC limit $n_\text{n,col}$ 
should account for most of the normal density. In the present unitary case, 
however, we expect both single-particle and collective excitations 
to be relevant.

Furthermore, Landau linked the normal densities to their statistic and 
their energy spectrum, see for instance Ref.~\cite{Fetter:1971uc}, so that 
in the present case the single-particle contribution to the 
normal density reads
\beq
n_{\text{n},\text{sp}} = \frac{2\beta}{3V} \sum_{\mathbf{k}} \frac{k^2}{m} 
\frac{e^{\beta \epsilon_\text{sp} (k)}}
{(e^{\beta \epsilon_\text{sp} (k)}+1)^2} \; ,
\eeq
whereas, concerning the contribution from the collective modes,
\beq
n_{\text{n},\text{col}} =\frac{\beta}{3V} \sum_{\mathbf{q}} \frac{q^2}{m} 
\frac{e^{\beta \omega_\text{col} (q)}}
{(e^{\beta \omega_\text{col} (q)}-1)^2} \; .
\eeq
It is then easy to derive the superfluid fraction 
\beq
{n_\text{s}\over n} = 1 - \Xi (x) \; , 
\label{fraction}  
\eeq
where the universal function $\Xi(x)$ is again a function of the scaled 
temperature $x \equiv T/T_F$ only, explicitly given by 
\beqa 
\Xi(x) &=& {2\over x} \int_0^{+\infty} 
{e^{{\tilde \epsilon}_\text{sp}(\eta)/x} \over 
(e^{{\tilde \epsilon}_\text{sp}(\eta)/x} + 1)^2} \  \eta^4 \mathrm{d} \eta \; 
\nonumber
\\
&+& {1\over x} \int_0^{+\infty} 
{e^{{\tilde \omega}_\text{col}(\eta)/x}\over 
(e^{{\tilde \omega}_\text{col}(\eta)/x} - 1)^2} \  \eta^4 \mathrm{d} \eta \; ,
\label{fraction1}
\eeqa
where we have converted sums to integrals. Finally, we stress that in the 
present model, the superfluid density defines the critical temperature 
$T_c$ via the condition $n_\text{s}=0$, and with our choice of parameters 
for the temperature-independent elementary excitation dispersions we find $T_c 
\approx 0.23 \ T_F$.
It must be pointed out that, while this estimation of the critical
temperature agrees with more refined approaches, such as the functional
GPF theory \cite{Hu:2007ez,Tempere:2012tt} or the NSR scheme \cite{Nozieres:1985zz},
it actually differs from the most recent experimental results,
placing it at $T_c/T_F \simeq 0.17$ \cite{Li:2022vn}. 
This shortcoming, shared among a range of different formalisms, is due to 
the fact the induced interaction is not taken into account 
\cite{yu-2009} according to the so-called Gorkov-Melik-Barkhudarov
theory \cite{gorkov}, which has been shown to provide the dominant
contribution on the BCS side and a relevant correction at unitarity.
The slight overestimation of our theoretical critical temperature  with respect to the
experimental one of Ref. \cite{Li:2022vn} 
does not appear plotting the physical quantities vs $T/T_c$.

\begin{figure*}[htbp]
\begin{center}
\includegraphics[width=\linewidth]{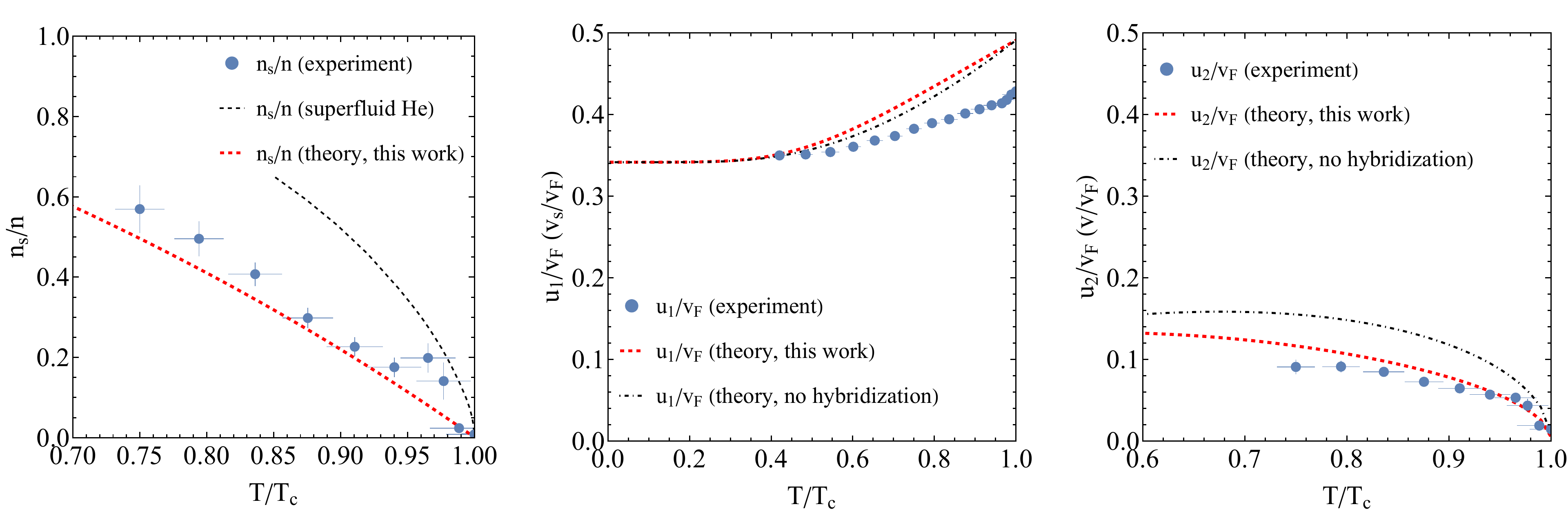}
\caption{Comparison between theory and experimental data in 
Ref.~\cite{Li:2022vn}  as a function of the dimensionless temperature 
$T/T_c$.
Left panel: superfluid fraction $n_\text{s}/n$.
Middle panel: adimensional first velocity $u_1/v_F$. 
Right panel: adimensional second sound velocity $u_2/v_F$.
In the central and right panels, we report the sound velocities computed in
absence of mixing, i.e. under the assumption that $c_{10} \approx c_T$. From the left panel,
we infer that, contrary to the $^{4}$He picture \cite{Landau:1980ue,stringari-book}, the equality 
between isothermal and adiabatic compressibilities reads a much worse agreement with the experimental
data, as evident from the behaviour of the second sound $u_2$ (right panel).}
\label{fig:one}
\end{center}
\end{figure*}

In the left panel of Fig. \ref{fig:one}, we report the
theoretically-derived superfluid fraction $n_\text{s}/n$ as a 
function of the dimensionless temperature $T/T_c$ (red dashed line), 
compared with experimental data \cite{Li:2022vn} for the unitary Fermi gas (blue dots), showing remarkable agreement;
as a reference, we also plot the critical-exponent 
behaviour observed in superfluid He (black dashed line). 

\section{First sound, second sound and sound mixing}

According to Landau \cite{Landau:1941vk,Khalatnikov:2000} a local
perturbation excites two wave-like modes – the first and the
second sound – which propagate with velocities $u_1$ and $u_2$.
These velocities are determined by the positive solutions
of the algebraic biquadratic equation (see also \cite{Tononi:2021wq})
\beq 
u^4 - (c_{10}^2+c_{20}^2) u^2 + c_T^2 c_{20}^2 = 0 \; ,  
\label{biquad}
\eeq
where 
\beq 
c_{10} = \sqrt{{1\over m} \left({\partial P 
\over \partial n}\right)_{\bar{S},V} } 
=  v_F \sqrt{ 
{5\over 9} \Phi(x) - 
{5\over 9} {T\over T_F} \Phi'(x)  } \; 
\label{c10}
\eeq
is the adiabatic sound velocity with $\bar{S}=S/N$ 
the entropy per particle, 
\beq 
c_{20} = \sqrt{{1\over m} {{\bar S}^2 \over 
\left({\partial {\bar S}\over \partial T}\right)_{N,V}} 
{n_\text{s}\over n_n} } =  
v_F \sqrt{ -{1\over 2} {\Phi'(x)^2 \over 
\Phi''(x)} 
{1-\Xi (x) \over \Xi (x)}} \;  
\label{c20}
\eeq
is the entropic sound velocity, and 
\beqa 
c_T &=& \sqrt{{1\over m} \left({\partial P 
\over \partial n}\right)_{T,V} } 
=
\nonumber 
\\
&=& v_F \sqrt{ {5\over 9}\Big( \Phi(x) - 
{T\over T_F}\Phi'(x) \Big) 
+ {2\over 9} x^2 
\Phi''(x)}
\label{cT}
\eeqa
is the isothermal sound velocity. It is immediate to find that 
for $T\to 0$ one has
\beqa
c_{10}\to c_B &=& v_F\sqrt{\xi/3} \\
c_{20}\to c_B/\sqrt{3} &=& v_F\sqrt{\xi}/3 \\
c_T\to c_B &=& v_F\sqrt{\xi/3}
\eeqa
The first sound $u_1$ is the largest of the two positive roots 
of Eq.~(\ref{biquad}) while the second sound $u_2$ is the 
smallest positive one. Thus 
\beq 
u_{1,2} = \sqrt{ {c_{10}^2 + c_{20}^2\over 2} 
\pm \sqrt{ \left( {c_{10}^2 + c_{20}^2\over 2}\right)^2 
- c_{20}^2 c_T^2}} \; . 
\label{u1eu2}
\eeq

We now compare our theory with the experimental data for the sound velocities from 
Ref.~\cite{Li:2022vn}. In particular, in the middle panel of Fig. \ref{fig:one} 
we plot the theoretically-calculated dimensionless first sound velocity $u_1/v_F$ as a function of the dimensionless 
temperature $T/T_c$ (red dashed line), comparing it with the experimental data \cite{Li:2022vn} (blue dots) 
showing quite good agreement with our theory. In the same panel we also plot the first sound calculated neglecting mode mixing,
i.e.~under the assumption that $c_T \approx c_{10}$ (black thin dashed-dotted line). 
In the right panel of Fig. \ref{fig:one} we plot the theoretically-derived 
dimensionless second sound velocity $u_2/v_F$ (red dashed line), compared with experimental 
data \cite{Li:2022vn} for the second sound velocity $u_2/v_F$ (blue dots).
In the same panel we also plot the dimensionless second sound $u_2/v_F$ calculated 
neglecting mode mixing (black thin dashed-dotted line).
As far as the second sound is concerned, our theory shows remarkable agreement with experimental data \cite{Li:2022vn}.
Importantly, this implies there is mixing between the first and second sound modes,
and that for the unitary Fermi gas it is wrong to assume an approximate 
equality of adiabatic and isothermal compressibilities. 

Concluding this Section, we stress that the Einstein-like relation
\beq
\frac{E}{N} = \frac{10}{9} m c_{10}^2
\label{eq:einstein}
\eeq
derived in Ref.~\cite{Patel:2020th} is automatically verified within our 
universal thermodynamic formalism, that naturally includes the 
scale-invariance of the unitary Fermi gas. 

\section{Response to a density perturbation}

\begin{figure}[htbp]
\begin{center}
\includegraphics[width=\linewidth]{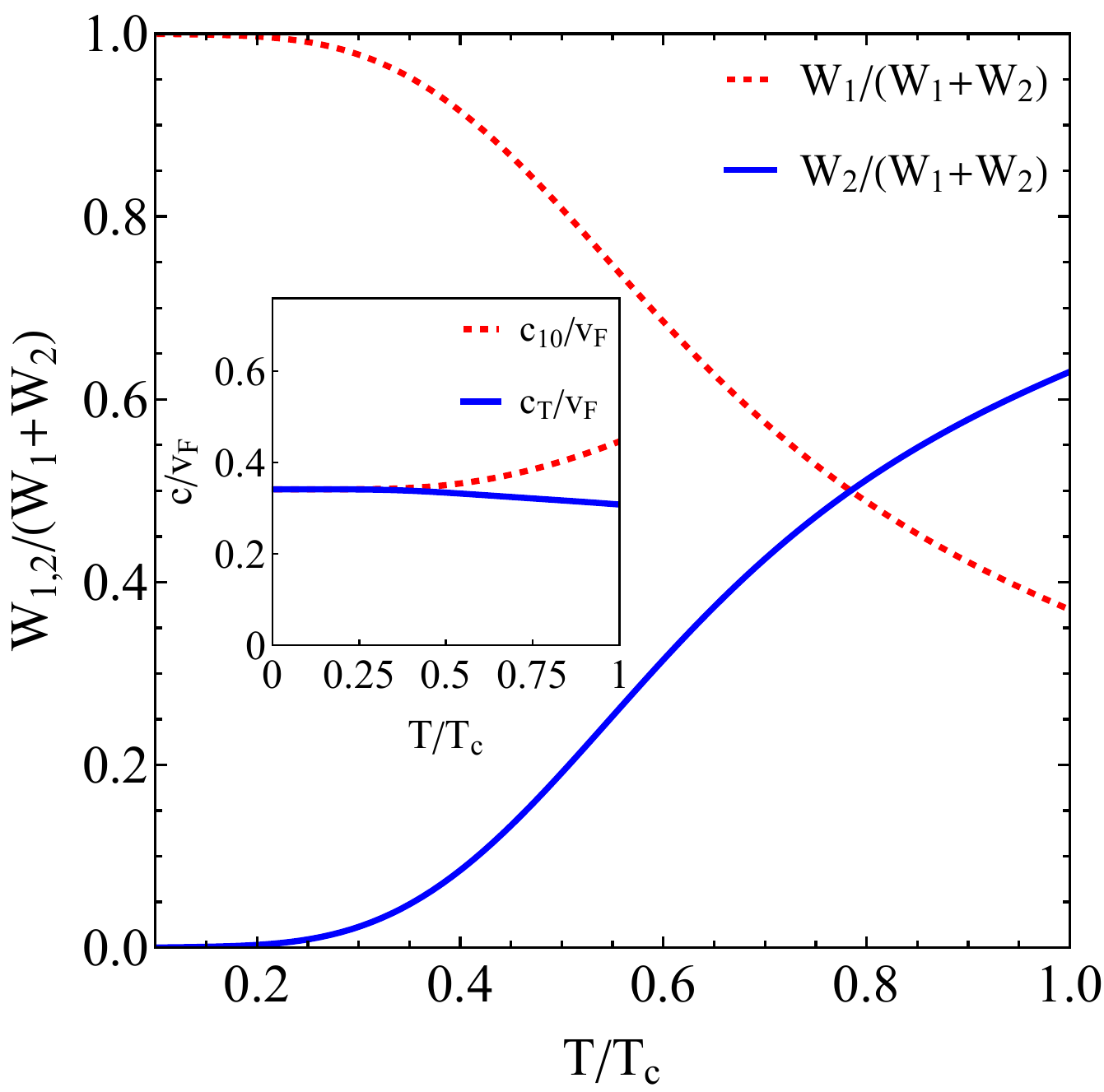}
\caption{\textit{Main Panel}. Contribution from the first (dashed red line) 
and second sound (solid blue line) to the amplitude of a density
response, as given by Eqs.~\eqref{w1} and \eqref{w2}, as a function of the 
scaled temperature $T/T_c$. \textit{Inset}. Adiabatic and entropic sound velocities,
$c_{10}$ and $c_T$ respectively (cfr. Eqs. \eqref{c10} and ), as functions of the scaled 
temperature. The no-mixing condition $c_{10}\approx c_T$ (see main text) is fulfilled for
$T/T_c \lesssim 0.4$, exactly where $W_2$ becomes vanishingly small.}
\label{fig:two}
\end{center}
\end{figure}

In general, the knowledge of the first and second sound velocities
may not be sufficient to provide a reliable characterization of the experimentally-observed
modes. First of all, we stress that the situation is radically different from what is observed
in superfluid $^4$He, where the response in density and temperature 
is decoupled and first sound corresponds to a standard 
density waves (in-phase oscillations of the superfluid and 
normal components), and the second sound is understood as an 
entropy wave \cite{Landau:1980ue,Pitaevskii:2015ul}. 
The technical reason has to be traced back to the isothermal and 
adiabatic compressibilities being approximately equal such that 
$c_{10} \approx c_T$, cfr.~Eq.~\eqref{c10} and Eq.~\eqref{cT}. 
However, this assumption does not hold for a generic quantum fluid, so that, 
in principle, even a simple density-perturbing protocol may excite 
both modes. This is exactly the case for ultracold bosons, 
for which, in two spatial dimensions, second sound acts as a reliable 
diagnostic tool for the onset of the BKT transition \cite{Hilker:2021us}. Moving
to Fermi gases, the situation across the BCS-BEC crossover is significantly
more involved \cite{Hoffmann:2021wi}: while the experimental setups is 
certainly not comparable to Helium,
there have been cases where a density-perturbing protocol excited just a single
mode \cite{Bohlen:2020vy,Tononi:2021wq}. 

Therefore, besides the values of $u_1$ and $u_2$ in Eq.~\eqref{u1eu2}, 
in order to provide a more complete characterization of 
the experimental picture, we also have to consider the amplitudes modes $W_1$
and $W_2$ of the response to a density perturbation 
\cite{Hoffmann:2021wi,Tononi:2021wq,stringari-book}, i.e. 
\begin{equation}
\delta \rho(x,t) = W_1 \delta\rho_1(x\pm u_1 t) + W_2\delta\rho_2(x\pm u_2 t)
\label{propagating perturbation}
\end{equation}
where
\begin{equation}
\frac{W_1}{W_1 + W_2}  = \frac{(u_1^2 - c_{20}^2)\, u_2^2}{(u_1^2 - u_2^2)
\, c^2_{20}} 
\label{w1}
\end{equation}
and
\begin{equation}
\frac{W_2}{W_1 + W_2} = \frac{(c_{20}^2 - u_2^2)\, u_1^2}{(u_1^2 - u_2^2)
\, c^2_{20}} \;.
\label{w2} 
\end{equation}

In Fig. \ref{fig:two} we report the behaviour of the relative amplitude 
contributions as a function of the temperature. Remarkably, we observe that
in the ultralow-$T$ regime a density probe actually excites only the first sound, since the 
amplitude of the $u_2$-mode vanishes as $T\rightarrow 0$. 
It is important to notice that, under the no-mixing condition $c_{10} \approx c_T$, 
Eqs. \eqref{w1} and \eqref{w2} read $W_1 = 1$ and $W_2 = 0$. Thus, this implies that
mode mixing is extremely reduced deeply below the critical temperature, as confirmed
by the inset in Fig. \ref{fig:two}, showing the no-mixing condition fulfilled at
$T \lesssim 0.4 \,T_c$. Moving closer to the transition, 
our theoretical model predicts that the balance between $W_1$ and $W_2$ should tip over around $T/T_c \simeq 0.8$, where the second-sound mode becomes the dominant one.
This means that, while in principle a density perturbation can excite
both modes, at $T\rightarrow 0$ (i.e.~deeply into the superfluid regime),
the amplitude corresponding to $u_2$ is vanishingly small and actually 
undetectable. The situation is overturned moving closer
to the critical temperature, where the superfluid susceptibility is much 
higher and both modes can be simultaneously excited with comparable amplitudes.

\section{Conclusions}

In this paper we have shown that a simple description in terms of
temperature-independent elementary 
excitations is able to reproduce many properties of the unitary Fermi gas: 
in particular we have reproduced the recently-measured 
superfluid fraction near the critical point \cite{Li:2022vn} and, 
after properly accounting for mixing between sounds modes, 
also the first and second sound velocities.
We have found that, contrary to liquid helium, 
near the critical temperature the first and second sound 
of the the unitary Fermi gas cannot be interpreted as a 
pure pressure-density wave and a pure entropy-temperature wave, respectively. 
We have also analyzed the density response to an external perturbation,
our investigation showing that at very low temperatures
the mixing of pressure-density and entropy-temperature oscillations
is absent, whereas approaching $T_c$ a density probe will excite both sounds.
Finally, we stress that Ref.~\cite{Li:2022vn} reports a measurement of
the sound diffusion from which they derive the viscosity-entropy ratio. 
Adopting the analysis developed in Refs.~\cite{How:2010uq,Salasnich:2011tj} our 
calculated viscosity-entropy ratio is about three times smaller than 
the one of Ref.~\cite{Li:2022vn} but, however, in good agreement
with previous experimental determinations \cite{Kinast:2004ug,Luo:2009uu,Cao:2011uu,Schafer:2012tx}.

\begin{acknowledgments}
The authors acknowledge stimulating discussions with T.~Enss. This work is supported by 
the Deutsche Forschungsgemeinschaft 
(DFG, German Research Foundation) under Germany's Excellence Strategy 
EXC2181/1-390900948 (the Heidelberg STRUCTURES Excellence Cluster).
\end{acknowledgments}

\end{document}